# Mining Public Opinion about Economic Issues: Twitter and the U.S. Presidential Election

Amir Karami
School of Library and Information Science, University of South Carolina, Columbia, SC, USA
karami@sc.edu

London S. Bennett
University of South Carolina Honors College, Columbia, SC, USA

lsb@email.sc.edu

Xiaoyun He
Department of Information Systems, Auburn University at Montgomery, Montgomery, AL, USA
xhe@aum.edu

**Amir Karami** is an Assistant Professor in the School of Library and Information Science and a Faculty Associate at the Arnold School of Public Health at the University of South Carolina. His research interests are text mining, computational social science, and medical/health informatics. He is currently working on developing fuzzy text mining techniques and their applications in medical, health, and social science. Previously, he was a Data Science Consultant at Washington DC area.

**London S. Bennett** is Political Science major in the University of South Carolina Honors College. Her work lies at the intersection of social computing, political science, psychology, and data science. She has also internship and study abroad experiences in legal and sociopolitical research. She has received multiple grants and awards for her scholarly pursuits.

**Xiaoyun He** is an Assistant Professor in Information Systems at the College of Business, Auburn University at Montgomery. She received her PhD from Rutgers University in Management with a concentration in Information Technology. Dr. He's research examines both technical and strategic aspects pertaining to information technology and management. Her research has led to a number of publications in peer-reviewed journals and conference proceedings.

## Abstract

Opinion polls have been the bridge between public opinion and politicians in elections. However, developing surveys to disclose people's feedback with respect to economic issues is limited, expensive, and time-consuming. In recent years, social media such as Twitter has enabled people to share their opinions regarding elections. Social media has provided a platform for collecting a large amount of social media data. This paper proposes a computational public opinion mining approach to explore the discussion of economic issues in social media during an election. Current related studies use text mining methods independently for election analysis and election prediction; this research combines two text mining methods: sentiment analysis and topic modeling. The proposed approach has effectively been deployed on millions of tweets to analyze economic concerns of people during the 2012 US presidential election.

## INTRODUCTION

Opinion polls and surveys are the bridge between public opinion and politicians and have played an important role in elections. An "opinion poll is a type of survey or inquiry designed to measure the public's views regarding a particular topic or series of topics" (Gallup World Poll, 2007). Different methods, including face-to-face interviews, phone interviews, and surveys sent by mail or email or available online have been used to gather and learn about people's thoughts about main issues (Cowling, 2015). There are two main approaches for mining public political opinions: qualitative and quantitative. Quantitative methods, such as polls, provide more data than the qualitative methods, such as interviews (Macreadie, 2011). Opinion polls have the power of influencing the views of people and parties during an election (Gallup World Poll, 2007). This power encourages political campaigns to track polls and surveys for possible changes in public relation strategies.

Among different technologies, social media plays the role of a big focus group in providing feedback during an election cycle (Macreadie, 2011). There are 2.3 billion active social media users, and Facebook Messenger and Whatsapp generate 60 billion messages per day (Smit, 2016). Among the popular social media, Twitter data has increasingly been used in a variety of contexts and has created interest for researchers. With 500 million short messages per day, Twitter opens a new opportunity for performing analysis on a huge amount of publicly available data. The use of Twitter for political campaigns is increasingly commonplace, from major national contests to regional and local elections in the United States, and with 80 million Twitter users (Stein, 2017).

Before 2016, the 2012 presidential campaigns of Barack Obama and Mitt Romney represented the most data-driven election cycle in history (Stromer-Galley, 2014). The Obama and Romney campaigns spent $52 million and $26 million on advertising in modern social media, respectively (Mazmanian, 2012). In that election, 40% of U.S.

adults engaged politically with social media, 38% of social media users shared and followed political news, and 20% of the users followed politicians on social media (Rainie et al. 2012).

People share their feelings and opinions on Twitter on such a large scale that it can be used as a valuable, publicly available resource for both academia and industry. Unlike traditional surveys, collecting and analyzing Twitter data is a cost-effective way to survey a large number of participants in a short period of time. Previous studies have used different analytical methods, such as finding high frequency words and applying text mining methods for election prediction and analysis; however, those studies have not considered the combination of advanced text mining methods.

The economy is the most important concern to voters in elections judging by the amount of positive and negative feedback to candidates' economic policies (Saad, 2012). The importance of economic issues to voters indicates that there is a need to disclose public opinion with respect to those issues. While the previous studies are useful in analyzing social media, there is no research to identify key major economic topics.

In order to address the limitations of traditional surveys and current studies, this study proposes an economics-based opinion mining approach to analyze election-related tweets, to gather positive and negative economic feedback within them, and better understand public opinion on economic issues. The proposed approach applies a combination of sentiment analysis and topic modeling methods on millions of tweets during the 2012 U.S. presidential election to disclose overall economic public opinion with respect to the candidates, Barack Obama and Mitt Romney.

## RELATED WORK

Among social media, Twitter is the most popular social media for researchers because it is very convenient to collect the activity of users. Based on this feature, a wide range of studies have been developed, from business (Mishne & Glance, 2006) to public response to public health (Karami & Shaw, 2017; Karami et al., 2018). In this section, we review the political studies using Twitter data for two purposes: election prediction and election analysis.

### Election Prediction

Previous studies analyzed Twitter data for election prediction using two approaches: linguistics analysis and mixed methods.

#### Linguistics Analysis

Different studies used the frequency of tweets to determine the popularity of candidates (Tumasjan et al., 2010; Gaurav et al., 2013; Boutet et al., 2012). A similar method was developed based on the frequency of tweets that mentioned names of political parties, political candidates, and contested constituencies to predict the 2011 Singapore general election (Skoric et al., 2012).

Among computational linguistics methods, sentiment analysis is one of the popular methods. There are two main approaches for sentiment analysis: the learning-based approach and lexicon-based approach (Khan et al., 2015). The first approach uses machine learning methods to build classifiers on frequencies of words in labeled data (Taboada et al., 2011). The second approach uses a dictionary of sentiment terms to disclose sentiment in a corpus based on the frequency of predefined positive and negative words in a document (Medhat et al., 2014). If the number of positive words is more than the number of negative words in a document, the overall feeling is assumed to be positive and vice versa.

Sentiment analysis was applied on tweets to compute a sentiment score and predict the result of an election in both offline mood (Chung & Mustafaraj, 2011) and real-time mood (Wang et al., 2012). Sentiment analysis was also combined with other methods, such as tweet counting, on the 2011 Dutch senate election tweets (Sang & Bos, 2012), and volume-based measures on the 2011 Irish general election tweets (Bermingham & Smeaton, 2011).

### *Mixed Methods*

Different mixed methods using multiple features and methods have been used for election prediction, such as combining sentiment analysis and message query–based retrieval (Balasubramanyanet et al. 2010), combining text mining and classification (Jahanbakhsh & Moon, 2014), and combining the frequency of tweets and retweets (Borondo et al., 2012).

### **Election Analysis**

There are some studies at the macro level, such as analyzing the social media strategy of candidates during an election (LaMarre & Suzuki-Lambrecht, 2013), examining the perceived link between social media and public opinion (Anstead & O'Loughlin, 2015), and using Twitter to disclose the possible relation between candidate salience and the candidates' level of engagement in Twitter (Hong & Nadler, 2012).

One of the advanced text mining methods that has been used for election analysis is topic modeling. Topic modeling is a computational technique used to group related words. For example, "tax," "plan," and "company" can be assigned into a cluster that is understood as a tax issue theme. Topic models find latent hidden structure in corpora and have been used for a wide range of applications during the last decade (Boyd-Graber et al., 2014). These models have been shown to be very effective for tasks such as summarization, information retrieval, and image labeling (Boyd-Graber et al., 2014).

Current political studies use sentiment analysis and text mining methods separately to find frequency of tweets with particular hashtags and positive and negative tweets (Rill et al., 2014), while others applied word frequency to find high frequency words for sentiment analysis for election prediction (Wang et al., 2012).

One study has applied topic modeling and sentiment analysis approaches separately for two different purposes (Jahanbakhsh & Moon, 2014). The first purpose was to find five common topics on 30 thousand tweets each over three days. The second goal was to develop a new sentiment analysis technique for election prediction. This research suffers from two problems. The first problem is related to the data analysis approach. Although this research collected 39 million tweets, only a sample of ~400 thousand tweets was selected for data analysis. The second problem is that the number of negative tweets was ignored for the evaluation section.

Our approach goes beyond simply analyzing frequency of words and hashtags to find semantic structure in tweets. This research applies a combination of sentiment analysis and topic modeling methods for more than 300 topics to explore and reveal the discussed positive and negative economic feedback in millions of tweets.

## METHODOLOGY AND RESULTS

This paper proposes an economic-based public opinion mining approach with four components: data collection, sentiment analysis, topic discovery, and analysis. Then we apply our approach on the tweets regarding the 2012 U.S. presidential election to detect and analyze negative and positive feedback with respect to main economic issues.

### Data

Twitter data can be collected with APIs (Application Programming Interfaces) (Twitter, 2017). APIs collect different forms of Twitter data for a user such as tweets, number of followers, and favorite tweets (Mejova et al., 2015). To access a large number of tweets, some related terms are needed to retrieve the relevant ones. This step comes with a data cleaning step to remove stopwords, such as "the," that do not have any semantic value. The output of this step are the tweets related to each of the candidates (Fig. 1).

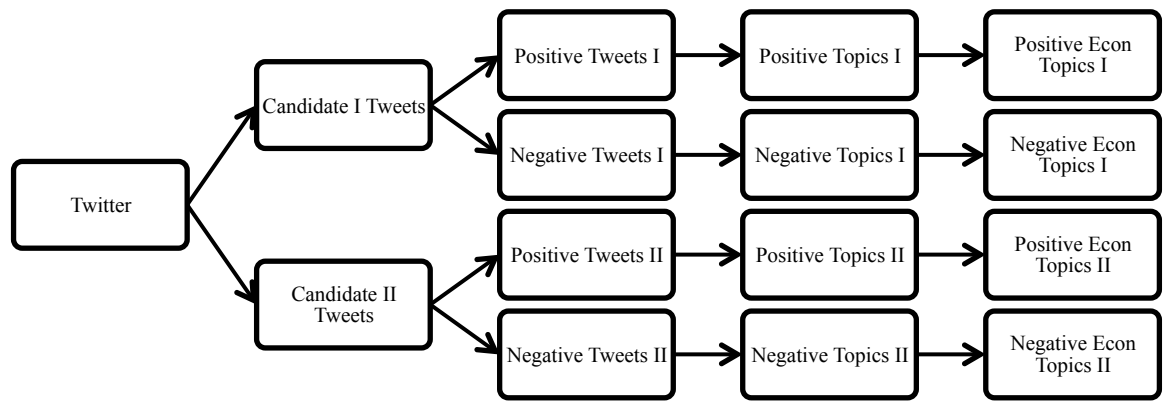

*Figure 1: Methodology Steps*

The data for this research was collected from September 29, 2012, to November 16, 2012, using the queries in Table1. This dataset has 24 million tweets, including the queries, related to the candidates for president, Barack Obama and Mitt Romney.

Table 1: Queries for Filtering Tweets

| Candidate | Queries |
|---|---|
| Barack Obama | barack obama |
| | @barackobama |
| | #barackobama |
| | #obama |
| Mitt Romney | mitt romney |
| | @mittromney |
| | #mittromney |
| | #romney |

**Sentiment Analysis**

The general approach to disclosing subjectivity and polarity from text data is sentiment analysis (Karami et al., 2018). Lexicon-based and learning-based approaches can be used for this step. The first approach uses machine learning classifiers when there is prior knowledge about data categories. In this case, a sample of the data is first labeled by human raters by assigning spam and non-spam labels to a sample of emails (Karami et al., 2018). The second approach, a cost-effective one, finds the frequency of a predefined dictionary of positive and negative terms to disclose sentiment in the data when there is no prior knowledge about its categories (Karami et al., 2018). We did not have any prior knowledge about the categories of the tweets in this research; therefore, we applied the second approach to find positive, negative, and neutral tweets. The collected tweets in the first step are filtered into positive and negative tweets in this step (Fig. 1).

In this study, we use Linguistic Inquiry and Word Count (LIWC) that is a lexicon-based tool for sentiment analysis (Pennebaker et al., 2007). LIWC is a text analysis tool to analyze linguistics features in a corpus. This tool has good sensitivity value, specificity value, and English proficiency measure (Golder & Macy, 2011; Karami & Zhou 2014a, 2014b, 2015). We filter the data to positive and negative tweets with respect to each candidate (Table 2). This analysis shows that Obama has the advantage based on the difference between positive and negative tweets. It would be worthwhile to explore the use of neutral tweets in future research.

Table 2: Sentiment Analysis Statistics

| Candidate | Sentiment | #Tweets |
|---|---|---|
| Barack Obama | Positive | 4,549,496 |
| | Negative | 3,075,592 |
| Mitt Romney | Positive | 2,773,933 |
| | Negative | 2,396,873 |

**Topic Discovery and Analysis**

Next we need to detect topics among the pool of the negative and positive tweets. Among different topic models, latent Dirichelt allocation (LDA) is the most popular effective topic model (Karami et al., n.d., 2015b). This model assumes that each word can be assigned to the topics in a corpus with a different degree of membership (Karami et al., 2015a). For example, LDA shows the relations between topics and words and discloses the themes, including Job in topic 1, Tax in topic 2, and Healthcare in topic 3 (Fig. 2). LDA helps us to find the topics in the positive and negative tweets (Fig. 1).

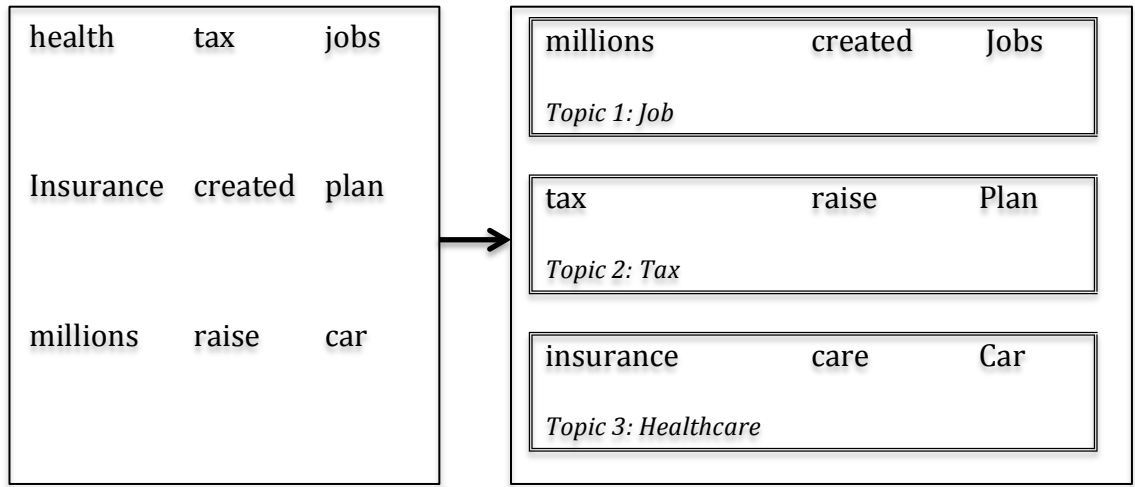

Figure 2: LDA Example

People posted tweets about different issues (topics) during the 2012 election, but the focus of this research is on the main economic issues including the *Economy in General, Job, Budget Deficit, Healthcare,* and *Tax* found in the published surveys (Pew Research Center, 2012; Gallup, 2012; Saad, 2012; Polling Report, 2012; Krieg, 2012; Gallup Editors, 2012). We consider these five issues in the detected positive and negative topics for each candidate (Fig. 1). It should be noted that analyzing other issues such as foreign policy would be a fruitful direction for future work.

We believe that the difference between the number of positive topics and the number of negative topics (DPNT) indicates the overall feedback status, including negative, positive, and neutral, for each of the economic issues per each candidate. Positive DPNT indicates that a candidate got more positive feedback than negative feedback with respect to an issue, and vice versa.

We applied the Mallet implementation of LDA (McCallum (2002)) with its default setting on negative and positive tweets to explore the topics for both of the candidates. We filtered the topics based on the five main economic issues, Economy in General, Job, Budget Deficit, Healthcare, and Tax. We detected a total of 325 economic topics including 70 positive topics and 62 negative topics associated with Obama, and 83 positive topics and 110 negative topics associated with Romney. Tables 3 and 4 show a sample of the detected topics. For example, we assigned the Job label to a topic containing "jobs," "created," "millions," "private," and "sector." This topic states that Obama created millions of jobs for the private sector.

Table 3: A Sample of Obama's Topics

| Economy | Job | Budget Deficit | Healthcare | Tax |
|---|---|---|---|---|
| good | jobs | debt | care | tax |
| economy | created | trillion | health | plan |
| markets | millions | deficit | obamacare | raise |
| grows | private | national | insurance | rich |
| succeed | sector | added | affordable | wealthy |

Table 4: A Sample of Romney's Topics

| Economy | Job | Budget Deficit | Healthcare | Tax |
|---|---|---|---|---|
| romney | romney | deficit | health | tax |
| bad | hurt | gov | insurance | plan |
| economy | thousands | debt | american | companies |
| idea | families | left | people | worse |
| policies | business | budget | americans | make |

Tables 5 and 6 show that Obama with a DPNT of +10 has the advantage over Romney who has a DPNT of -9 for the job issue. If both candidates get a negative DPNT for an issue, the candidate with higher DPNT value has the advantage. For instance, Obama with a DPNT of -5 has the advantage over Romney who has a DPNT of -10 for the tax issue. In total, Obama's DPNT is positive and Romney's DPNT is negative.

Table 5 shows that Obama was identified with more topics for the job issue. This output indicates that Job was the most important economic issue for Obama's followers on Twitter, followed by Economy in General, Tax, Healthcare, and Budget Deficit. Romney was also identified with more topics for the job issue followed by Tax, Economy in General, Healthcare, and Budget Deficit (Table 6). Obama has three positive DPNTs with the highest DPNT for the job issue and Romney has just one positive DPNT for the healthcare issue. In both cases, the tax issue is among the issues with the lowest DPNT. Although Obama has two negative DPNTs for the economy in general and tax issues, he has the advantage on all the economic issues based on DPNT value. The biggest and smallest gaps between the candidates are for the job and budget deficit issues, respectively.

Table 5: Obama DPNT Analysis

| Issues | Economy in General | Job | Budget Deficit | Healthcare | Tax |
|---|---|---|---|---|---|
| *#Positive Topics for Obama* | 13 | 34 | 4 | 11 | 8 |
| *#Negative Topics for Obama* | 18 | 24 | 3 | 4 | 13 |
| DPNT | -5 | +10 | +1 | +7 | -5 |

Table 6: Romney DPNT Analysis

| Issues | Economy in General | Job | Budget Deficit | Healthcare | Tax |
|---|---|---|---|---|---|
| *#Positive Topics for Romney* | 19 | 22 | 3 | 18 | 21 |
| *#Negative Topics for Romney* | 25 | 31 | 9 | 14 | 31 |
| DPNT | -6 | -9 | -6 | +4 | -10 |

## EVALUATION

We found two similar surveys with respect to the economic issues. The first survey was developed by Gallup (Saad, 2012); however, Gallup stated that flawed methods were used for the surveys of the 2012 presidential election (Moore, 2013). The second survey was conducted by the Pew Research Center which correctly predicted the winner. Therefore, we compare our results with the Pew survey (Pew Research Center, 2012).

Table 7 demonstrates the comparison of our results with the Pew survey. The second column in Table 7 shows the candidate with higher DPNT value advantage. For example, Obama's DPNT (-5) is higher than Romney's DPNT (-10) for the tax issue; therefore, we mention Obama in the second column of Table 7. The third column shows the candidate with advantage in the September 2012 Pew survey results (Pew Research Center, 2012). This table indicates that the Pew survey results are mostly (4 out of 5) in agreement with our analysis. Furthermore, the final election results show that Obama had a big victory with more than 3 million popular votes and a more than 120 electoral vote advantage over Romney (NBC News, 2012). In line with the final results, our analysis indicates that the winner had the advantage on the most important issues (economic issues) in the election.

Table 7: This Research vs the September 2012 Pew Survey

| Issue | Advantage in This Research | Advantage in the Pew Survey |
| --- | --- | --- |
| *Economy in General* | Obama | Obama |
| *Job* | Obama | Obama |
| *Budget Deficit* | Obama | Romney |
| *Healthcare* | Obama | Obama |
| *Tax* | Obama | Obama |

## CONCLUSION

One of the main concerns in an election is detecting and analyzing public feedback on important issues. Social media such as Twitter let millions of users share their opinions. This huge amount of data provides a great opportunity for public opinion mining.

This research has investigated public positive and negative economic feedback in election-related tweets. This paper analyzes millions of tweets systematically to provide new insight into social media public opinion mining and opens a new direction for future studies. This research proposes a computational approach using combination of sentiment analysis and topic modeling. This approach was applied on the tweets related to the 2012 U.S. presidential election to explore economic issues in positive and negative tweets. The proposed approach can find the distribution of negative and positive economic feedback in the related tweets for each candidate. The difference between the number of positive topics and the number of negative topics (DPNT) indicates the overall economic feedback for each candidate.

The results show that jobs and taxes were the most and the least important issues, respectively, for the followers of the two candidates. Although the overall ranking of the issues for each candidate is very close, DPNT values show Obama having the advantage on all the economic issues. An agreement of 80% between our results and a valid traditional survey demonstrates that the combination of sentiment analysis and topic modeling is a powerful approach for economics-based public opinion mining using Twitter data.

Computational content analysis of social media still needs improvement, and future research must also analyze noneconomic issues such as foreign policy. This research provides a basis for analyzing text data in social media and helps politicians, public opinion analysts, and social scientists better track main issues in both political and nonpolitical events. In the future, we will explore time and location variables to investigate the economic issues based on different time frames and locations.

**Acknowledgments**- This research is supported in part by the University of South Carolina Honors College Exploration Scholars Research Program Grant. All opinions, findings, conclusions, and recommendations in this paper are those of the authors and do not necessarily reflect the views of the funding agency. We would also like to thank Kazem Jahanbakhsh for sharing the Twitter dataset.

## REFERENCES

Anstead, N. & O'Loughlin, B. (2015), 'Social media analysis and public opinion: The 2010 uk general election', Journal of Computer-Mediated Communication 20(2), 204–220.

Balasubramanyan, R., Routledge, B. R. & Smith, N. A. (2010), 'From tweets to polls: Linking text sentiment to public opinion time series.', ICWSM 11 .

Bermingham, A. & Smeaton, A. F. (2011), 'On using twitter to monitor political sentiment and predict election results',Sentiment Analysis where AI meets Psychology (SAAIP) p. 2.

Borondo, J., Morales, A., Losada, J. C. & Benito, R. M. (2012), 'Characterizing and modeling an electoral campaign in the context of twitter: 2011 spanish presidential election as a case study', Chaos: an interdisciplinary journal of nonlinear science 22(2), 023138.

Boutet, A., Kim, H. & Yoneki, E. (2012), What's in your tweets? i know who you supported in the uk 2010 general election, in 'The International AAAI Conference on Weblogs and Social Media (ICWSM)'.

Boyd-Graber, J., Mimno, D. & Newman, D. (2014), 'Care and feeding of topic models: Problems, diagnostics, and improvements', Handbook of Mixed Membership Models and Their Applications.

Chung, J. & Mustafaraj, E. (2011), Can collective sentiment expressed on twitter predict political elections?, in 'Pro- ceedings of the Twenty-Fifth AAAI Conference on Artificial Intelligence', AAAI Press, pp. 1770–1771.

Cowling, D. (2015), 'How political polling shapes public opinion'. http://www.bbc.com/news/uk-31504146. Gallup (2012), 'Most Important Problem', http://www.gallup.com/poll/1675/most-important-problem.aspx.

Gallup Editors (2012), 'Romney 49%, Obama 48% in Gallup's Final Election Survey'. http://www.gallup.com/ poll/158519/romney-obama-gallup-final-election-survey.aspx.

Gallup World Poll (2007), 'What Is Public Opinion Polling and Why Is It Important?'. http://media.gallup.com/muslimwestfacts/PDF/PollingAndHowToUseItR1drevENG.pdf.

Gaurav, M., Srivastava, A., Kumar, A. & Miller, S. (2013), Leveraging candidate popularity on twitter to predict election outcome, in 'Proceedings of the 7th Workshop on Social Network Mining and Analysis', ACM, p. 7.

Golder, S. A. & Macy, M. W. (2011), 'Diurnal and seasonal mood vary with work, sleep, and daylength across diverse cultures', Science 333(6051), 1878–1881.

Hong, S. & Nadler, D. (2012), 'Which candidates do the public discuss online in an election campaign?: The use of social media by 2012 presidential candidates and its impact on candidate salience', Government Information Quarterly 29(4), 455–461.

Jahanbakhsh, K. & Moon, Y. (2014), 'The predictive power of social media: On the predictability of us presidential elections using twitter', arXiv preprint arXiv:1407.0622.

Karami, A., Dahl, A. A., Turner-McGrievy, G., Kharrazi, H. & Shaw, G. (2018), 'Characterizing diabetes, diet, exercise, and obesity comments on twitter', International Journal of Information Management 38(1), 1–6.

Karami, A., Gangopadhyay, A., Zhou, B. & Kharrazi, H. (2015a), Flatm: A fuzzy logic approach topic model for medical documents, in 'Annual Meeting of the North American Fuzzy Information Processing Society (NAFIPS)'.

Karami, A., Gangopadhyay, A., Zhou, B. & Kharrazi, H. (2015b), A fuzzy approach model for uncovering hidden latent semantic structure in medical text collections, in 'Proceedings of the iConference'.

Karami, A., Gangopadhyay, A., Zhou, B. & Kharrazi, H. (n.d.), 'Fuzzy approach topic discovery in health and medical corpora', International Journal of Fuzzy Systems pp. 1–12.

Karami, A. & Shaw, J. G. (2017), Computational content analysis of negative tweets for obesity, diet, diabetes, and exercise, in 'Annual Meeting of the Association for Information Science and Technology (ASIST)'.

Karami, A. & Zhou, B. (2015), Online review spam detection by new linguistic features, in 'iConference'.

Karami, A. & Zhou, L. (2014a), Exploiting latent content based features for the detection of static sms spams, in 'Annual Meeting of the Association for Information Science and Technology (ASIST)'.

Karami, A. & Zhou, L. (2014b), Improving static sms spam detection by using new content-based features, in 'Amer- icas Conference on Information Systems (AMCIS)'.

Khan, A. Z., Atique, M. & Thakare, V. (2015), 'Combining lexicon-based and learning-based methods for twitter sen- timent analysis', International Journal of Electronics, Communication and Soft Computing Science & Engineering (IJECSCSE) p. 89.

Krieg, G. J. (2012), 'Where Obama and Romney Stand on the Big Issues'. http://abcnews.go.com/Politics/ OTUS/obama-romney-stand-big-issues/story?id=17611080.

LaMarre, H. L. & Suzuki-Lambrecht, Y. (2013), 'Tweeting democracy? examining twitter as an online public relations strategy for congressional campaigns', Public Relations Review 39(4), 360–368.

Macreadie, R. (2011), Public Opinion Polls, Research Service, Parliamentary Library, Department of Parliamentary Services.
Mazmanian, A. (2012), 'Republicans Flame Romney's Digital Team'. http://mashable.com/2012/11/16/mitt- romney-digital-team/#kfWE35DzhOqh.

McCallum, A. K. (2002), 'MALLET: A Machine Learning for Language Toolkit.'. http://mallet.cs.umass. edu/topics.php.

Medhat, W., Hassan, A. & Korashy, H. (2014), 'Sentiment analysis algorithms and applications: A survey', Ain Shams Engineering Journal 5(4), 1093–1113.
Mejova, Y., Weber, I. & Macy, M. W. (2015), Twitter: a digital socioscope, Cambridge University Press.

Mishne, G., & Glance, N. S. (2006, March). Predicting Movie Sales from Blogger Sentiment. In AAAI Spring Symposium: Computational Approaches to Analyzing Weblogs (pp. 155-158).

Moore, M. T. (2013), 'Gallup identifies flaws in 2012 election polls'. https://www.usatoday.com/story/news/ politics/2013/06/04/gallup-poll-election-obama-romney/2388921/.

NBC News (2012), 'Presidential Election Results'. http://elections.nbcnews.com/ns/politics/2012/all/ president/#.WPwTSVPyuu4.

Pennebaker, J. W., Booth, R. J. & Francis, M. E. (2007), 'Linguistic inquiry and word count: Liwc [computer soft- ware]', Austin, TX: liwc. net .

Pew Research Center 10/24/12 (2012), 'Public Opinion on Foreign Policy'. http://www.people-press.org/2012/10/24/mitt-romney-barack-obama-on-foreign-policy/.

Pew Research Center September 9/19/12 (2012), 'Section 1: The Obama-Romney Matchup'. http://www.people-press.org/2012/09/19/section-1-the-obama-romney-matchup/.

Polling Report (2012), 'Problems and Priorities'. http://www.pollingreport.com/prioriti.htm.

Rainie, L., Smith, A., Schlozman, K. L., Brady, H. & Verba, S. (2012), 'Social media and political engagement', Pew Internet & American Life Project 19.

Rill, S., Reinel, D., Scheidt, J. & Zicari, R. V. (2014), 'Politwi: Early detection of emerging political topics on twitter and the impact on concept-level sentiment analysis', Knowledge-Based Systems 69, 24–33.

Saad, L. (2012), 'Economy Is Dominant Issue for Americans as Election Nears'. http://www.gallup.com/poll/ 158267/economy-dominant-issue-americans-election-nears.aspx.

Sang, E. T. K. & Bos, J. (2012), Predicting the 2011 dutch senate election results with twitter, in 'Proceedings of the Workshop on Semantic Analysis in Social Media', Association for Computational Linguistics, pp. 53–60.

Skoric, M., Poor, N., Achananuparp, P., Lim, E.-P. & Jiang, J. (2012), Tweets and votes: A study of the 2011 singapore general election, in '2012 45th Hawaii International Conference on System Science (HICSS)', IEEE, pp. 2583–2591.

Smit, K. (2016), 'Marketing: 96 Amazing Social Media Statistics and Facts'. https://www.brandwatch.com/ 2016/03/96-amazing-social-media-statistics-and-facts-for-2016.

Stein, J. (2017), 'Why snapchat's snappy', Time .

Stromer-Galley, J. (2014), Presidential campaigning in the Internet age, Oxford University Press.

Taboada, M., Brooke, J., Tofiloski, M., Voll, K. & Stede, M. (2011), 'Lexicon-based methods for sentiment analysis', Computational linguistics 37(2), 267–307.

Tumasjan, A., Sprenger, T. O., Sandner, P. G. & Welpe, I. M. (2010), Predicting elections with twitter: What 140 characters reveal about political sentiment., in 'ICWSM 10'.

Twitter (2017), 'Twitter Developer Documentation'. https://dev.twitter.com/docs.

Wang, H., Can, D., Kazemzadeh, A., Bar, F. & Narayanan, S. (2012), A system for real-time twitter sentiment analysis of 2012 us presidential election cycle, in 'Proceedings of the ACL 2012 System Demonstrations', Association for Computational Linguistics, pp. 115–120.